# Dark matter in the inner parts of barred galaxies: The data[⋆,⋆⋆]

I. Pérez[1,3], I. Márquez[2], K. Freeman[3], and R. Fux[3,4]

[1] Kapteyn Astronomical Institute, University of Groningen, Postbus 800, Groningen 9700 AV, The Netherlands
    e-mail: isa@astro.rug.nl
[2] Instituto de Astrofísica de Andalucía (CSIC), Apartado 3004, 18080 Granada, Spain
[3] RSAA, Mount Stromlo Observatory, Cotter road, Weston Creek, ACT 2611, Australia
[4] Geneva Observatory, Ch. des Maillettes 51, 1290 Sauverny, Switzerland

**Abstract.** This paper presents surface photometry ($B, V, I, J, H, K$) and $H_\alpha$ rotation curves of 27 isolated spiral galaxies. The final goal is to obtain the mass distribution of a sample of isolated spiral galaxies in order to model their gas kinematics. This is then compared to the observed rotation curve, to determine the necessity of a dark halo in the inner parts (Pérez et al. 2004). The azimuthally averaged radial surface brightness profiles and the integrated magnitudes obtained from ellipse fitting are given for each of the sample galaxies. The ellipse fitting technique applied to the light distribution also allowed us to obtain the size of the bar, and the inclination and position angle of the outer isophotes that allow the galaxy deprojection. Using these profiles, 1-D disk-bulge decomposition was performed to obtain the disk scale-length and the bulge effective radius for the different bands. Through the fitting of a parametric function to the observed rotation curve, the maximum rotational velocity and the corresponding radius was obtained. The correlation between the bulge and disk parameters is in agreement with previous studies (de Jong 1996a; Márquez & Moles 1999; Baggett et al. 1998). Regarding the Kormendy relation (Kormendy 1977), in agreement with de Jong, no correlation between the bulge effective radius and its surface brightness is found, possibly due to the small range of bulge magnitudes covered. We find a smaller scatter in the structural relations when compared to non-isolated samples in agreement with Márquez & Moles (1999). Finally, a correlation between the disk scale-length and the bar size is observed, possibly reflecting the rapid growth of a bar.

**Key words.** galaxies: fundamental parameters – galaxies: structure – galaxies: photometry – galaxies: kinematics and dynamics

## 1. Introduction

The coupling between the massive dark halo and the visible component in spiral galaxies leads to a degenerate problem. The velocity field of an axisymmetric galaxy does not allow to uniquely disentangle the contribution of both components. The "maximum disk" assumption requires that the stellar disk accounts for most of the rotational support of the galaxy in the inner parts. There is no consensus yet on whether maximum disks are preferred (Freeman 1992; Courteau & Rix 1999; Salucci & Persic 1999). The match to the overall shape of the rotation curve is often met just as well by pure halo or pure disk models. It is not even clear whether the Milky Way favours a maximum disk or not (Lewis & Freeman 1989; Fux & Martinet 1994; Sackett 1997; van der Kruit 1999).

To test whether the luminous mass in the inner parts of spiral galaxies can account for their observed gas kinematics or whether a more axisymmetric dark matter component is required, the inner gas dynamics of a sample of barred spiral galaxies has been modelled by running a 3-D composite $N$-body/hydrocode on the potential derived directly from the light distribution. This approach takes into account the non-axisymmetric structures, such as bars, that are very important in the dynamics of the inner region. These results are presented in a companion paper; Pérez et al. (2004), hereafter Paper I.

This paper presents the initial data set and sample characterization of a sample of isolated spiral galaxies from which the modelled galaxies presented in Paper I were chosen. This paper also presents the data analysis required to prepare the galaxies for the modelling of their stellar mass distribution prior to the dynamical simulations.

Since the final aim is to investigate the distribution of luminous matter and test whether an additional dark component is needed to explain the observed kinematics, a good tracer of the stellar mass distribution is needed. Near infrared (NIR) observations have been chosen because they are good tracers of the underlying stellar population. The light in the NIR bands is dominated by the main sequence low mass stars, although there could also be an important contribution from giant stars.

---

Furthermore, the NIR is less affected by internal extinction than the optical bands. Long slit observations were chosen as kinematic tracers. Integral field spectrographs offer an advantageous way of obtaining the kinematics of galaxies. However, long slit spectrographs have several features that make long-slit observations an interesting instrument for kinematic studies; it has more wavelength and spatial coverage than most 2-D spectrographs. The wavelength coverage is very useful to characterise the internal extinction of the galaxies and the large spatial coverage is useful to reach the regions well outside the bar region. Furthermore, very high signal-to-noise can be achieved for all the covered wavelengths. Different slit positions can help to understand the modelled 2-D velocity field.

To obtain reliable mass-to-light ratios through population synthesis, is necessary to obtain photometry in different bands. The available stellar libraries are more complete for the optical bands so together with the NIR observations, further *B*, *V* and *I* band observations were obtained. Extinction information can be obtained from the Balmer line ratios which can help to determine the opacity of the disk as a function of radius. This information can be obtained from the spectroscopic observations. The colour maps can also help to trace the dust absorption: the profiles are expected to redden with dust absorption. Since the galaxies in the sample are too faint in HI to obtain high resolution gas maps, the 2-D dust maps will help tracing the gas and will be compared to the simulated gas distribution (Paper I).

The selection of the sample will be described in Sect. 2. In Sect. 3 the observations and data reduction of the galaxies of the sample is explained. The methodology to prepare the images for the SPH modelling is presented in Sect. 4. The methodology and the correlation between the different parameters is presented in Sect. 5. Finally, the conclusions are given in Sect. 6

## 2. Sample selection

The properties of an initially unperturbed galaxy can be significantly modified by gravitational interaction with other galaxies. Such modifications, showing up either in the morphology or the dynamical properties, can be relatively mild (forming floculent like structures or tidal bridges) or so strong that they could lead to the complete disruption of the original system to form a different one, like an elliptical galaxy from the merging of two spiral galaxies (Barnes & Hernquist 1996).

To avoid environmental effects where dark halos could be merged or overlapped it is necessary to study a well defined sample of isolated galaxies. It is, of course, not a simple question to define an isolated galaxy. Here, we define an operational criteria to ensure that possible perturbation effects have been erased, assuming that the typical time scale for the perturbation effect decay is no longer than a few times $10^9$ years. For more detailed explanation about the selection process and isolation criteria see Márquez et al. (1999, 2000).

We defined isolated systems as those without a companion within a projected radius of 0.5 Mpc or within the volume given by $\Delta(cz) \leq 500$ km s$^{-1}$. In this way, one ensures that no encounter occurred between the galaxies in a time scale of the order of $10^9$ yr. The initial sample was chosen as follows (see Fig. 1): all the isolated spiral galaxies in the volume given by $cz \leq 6000$ km s$^{-1}$ with galactic longitude(b) $\geq 40°$, to avoid being affected much by the extinction of our galaxy, and those with inclination between 30° and 70°. These inclination angles were chosen to ensure that good measurement of the rotation curve was done; also, not too inclined galaxies were chosen to avoid internal extinction and to simplify the deprojection process.

A preliminary list was created using the Third Reference Catalogue of bright galaxies (RC3), de Vaucouleurs et al. (1991), for the southern hemisphere. 2100 galaxies were chosen to be southern ($\delta < 15°$) and volume confined (see above). From these, 293 were chosen to be isolated, using the criteria described above. However, non-classified objects could have still been found close to these galaxies. Therefore, all candidates were checked for companions with unknown redshifts using the Digital Sky Survey (DSS). Good quality data was obtained for 27 galaxies out of the final sample of 35 isolated galaxies. Up to this date, 5 barred galaxies have been modelled using hydrodynamical simulations (Paper I).

### 2.1. Characterisation of the sample

Figure 2 shows the number of galaxies as a function of morphological type. The morphological types were taken from the RC3 catalog and are based on the *B*-band; no attempt has been made to compare the morphologies in the different observed bands (apart from the presence of bars). Table 3 presents the main properties of the sample galaxies. From the sample of 35 galaxies, 18 galaxies (51%) present a bar (RC3 classification) in the optical band. From the final 27 galaxies, 8 have no information about the presence of a bar and 3 are classified as non-barred. After the ellipse fitting analysis (Sect. 5.3) and visual check, from the 8 galaxies with no bar information 3 showed clearly in the NIR or the *I*-band a bar morphology (UGC 10130, UGC 4861 and UGC 1553). From the 3 galaxies previously classified as non-barred, two showed bar morphologies in the NIR (NGC 6014 and ESO417-G006). With respect to AGN presence, 4 (11%) galaxies present an active nucleus. These numbers are found in larger samples. Due to the small number statistics these numbers are just indicative and are only useful to characterise the sample. One of the galaxies, NGC 6014, has been reclassified as a Sy2 galaxy using the Osterbrok diagnosis diagrams (Osterbrock 1989), only the [OII] emission line is not covered in the spectral range observed. The final sample of 27 galaxies does not differ in distribution of parameters to that of the initial 35 galaxies (all the AGN, except NGC 6014, are part of the final sample).

## 3. Observations and reduction

### 3.1. Long-slit observations

The rotation curves were obtained from optical spectroscopic (H$\alpha$) observations along several position angles using the Cassegrain Boller & Chivens spectrograph on the 1.52 m telescope at La Silla Observatory and the Double Beam Spectrograph (DBS) on the 2.3 m telescope at Siding

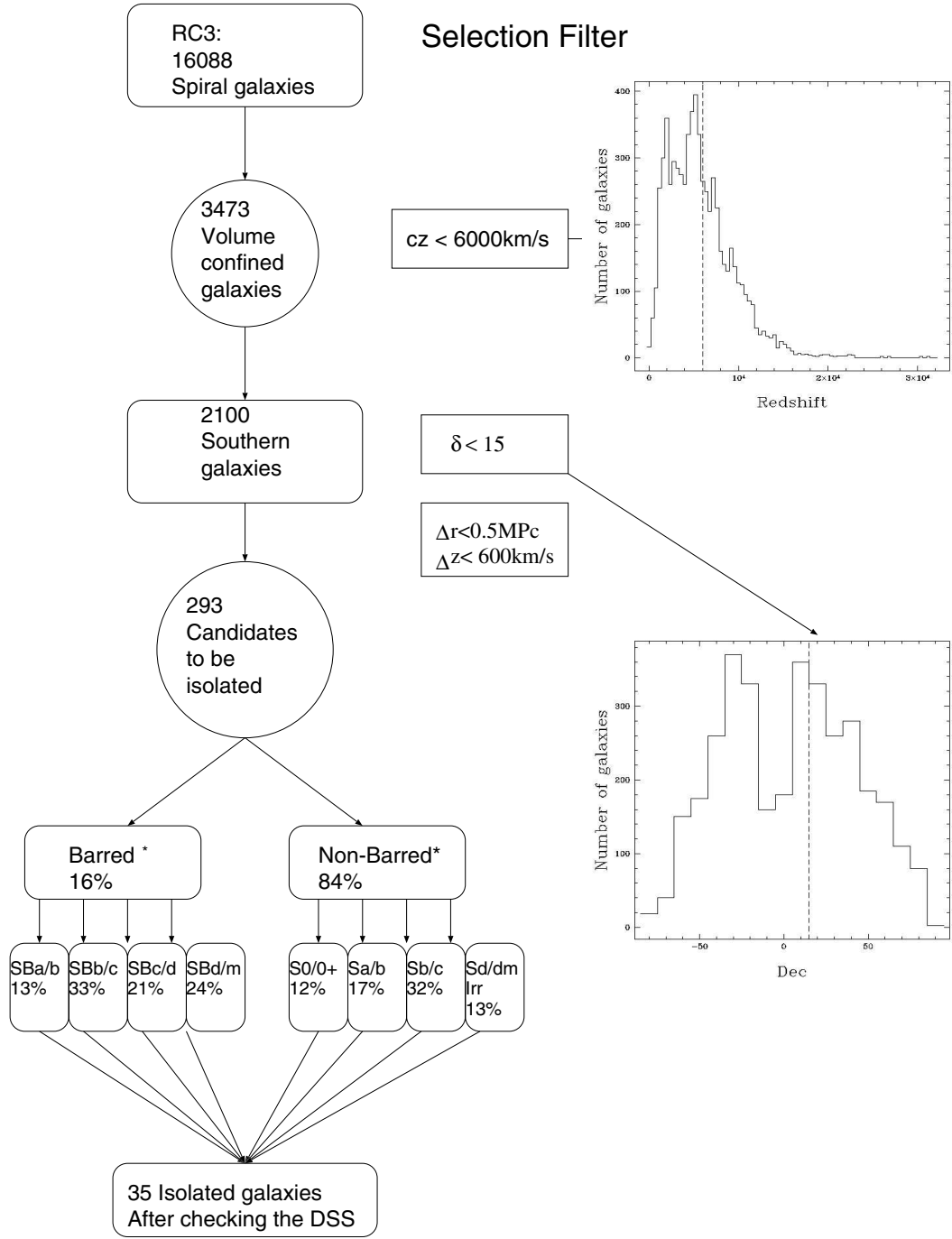

**Fig. 1.** Flow diagram showing the selection criteria.

Spring Observatory (SSO). At La Silla, the detector was a $2K \times 2K$ Loral-Lesser CCD with a pixel scale of 0.82 arcsec. The achived dispersion was 1.0 Å pixel$^{-1}$ using grating #26, with a wavelength coverage of $\lambda$ 4900–5700 Å. The detector at SSO was a SITe $1752 \times 532$ CCD. The gratings used were the 1200B and the 1200R for the blue and red arms, respectively. This set-up gives a dispersion of 1.13 Å pixel$^{-1}$ for the blue arm and 1.09 Å pixel$^{-1}$ for the red arm in the wavelength interval from 4315–5283 Å and 6020–6976 Å ; respectively, giving a dispersion resolution $\Delta\sigma \approx 45$ km s$^{-1}$ . The slit was in both cases visually aligned with the galaxy nucleus at major and minor axis (RC3) and the slit width was set to 1.5 arcsec on the sky, giving a projected width of 2.2 pixels. The average seing was 1.2 arcsec.

A comparison lamp exposure was obtained before and after each exposure to wavelength calibrate the frames. Spectrophotometric standards were observed with a slit of 10 arcsec. Total integration times for the galaxies varied from 1 to 3 h. In the case of the DBS the exposures had to be shorter than 20 minutes due to flexure of the instrument. The observations were split in three or more exposures per galaxy to eliminate cosmic rays from the frames. The observing log is presented in Table 4.

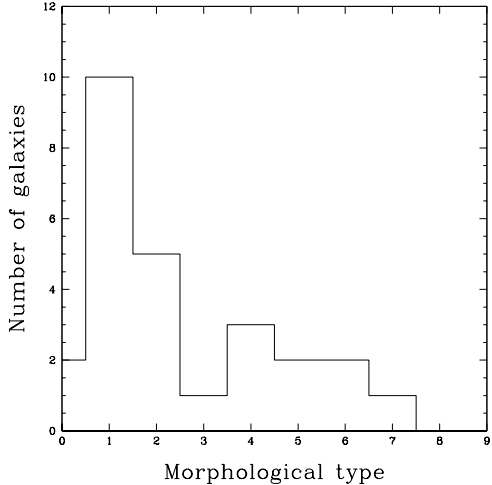

**Fig. 2.** Morphological type distribution for the final sample of 27 galaxies.

**Table 1.** Calibration coefficients of the optical observations.

| Band | Zero-point | Colour coeff. | Airmass coeff. |
|---|---|---|---|
| | | June 1999 LCO | |
| B | −21.19 ± 0.050 | 0.060 ± 0.020 | −0.250 ± 0.030 |
| V | −22.44 ± 0.190 | −0.080 ± 0.010 | −0.230 ± 0.150 |
| I | −21.67 ± 0.040 | 0.014 ± 0.010 | −0.070 ± 0.030 |
| | | January 1999 LCO | |
| B | −22.19 ± 0.050 | 0.030 ± 0.011 | −0.210 ± 0.040 |
| V | −22.25 ± 0.040 | −0.050 ± 0.010 | −0.090 ± 0.030 |
| I | −21.65 ± 0.044 | 0.010 ± 0.010 | 0.008 ± 0.035 |
| | | June 2001 SSO | |
| B | −22.40 ± 0.037 | −0.037 ± 0.004 | −0.214 ± 0.032 |
| V | −22.70 ± 0.068 | 0.008 ± 0.008 | −0.111 ± 0.059 |
| I | −22.14 ± 0.140 | −0.025 ± 0.017 | −0.188 ± 0.122 |

All the spectra obtained were reduced using standard IRAF routines. Overscan and bias was subtracted from all the frames. No dark subtraction was done due to the low dark current of the chips used (1 e$^{-}$ pix$^{-1}$ h$^{-1}$). At the 1.52 m telescope, dome and sky flats were acquired in order to correct properly for slit width variation. Flatfielding correction was achieved to the 2% level. Geometrical corrections were applied to the frames in order to correct for any misalignment. The relatively small, compared to the slit size, size of the galaxies allowed for a correct sky subtraction for each frame prior to combining. A first order polynomial was fit along the spatial direction and then subtracted from the frames. In most of the cases, all the sky lines were well subtracted, but for some of the brightest lines some residuals persisted and had to be masked manually. The data reduction for the DBS spectra was carried out in a similar way, but due to the double beam nature of the spectrograph, care had to be taken to deconvolve the dichroic using spectrophotometric standards.

Ionised gas rotation curves were obtained by fitting a Gaussian to the H$\alpha$ and the [NII] ($\lambda$6583.4 Å) emission lines using the MIDAS package ALICE. These lines were clearly detected in all the galaxies observed. The gas curves derived independently from H$\alpha$ and [NII] are in good agreement at all radii. The H$\alpha$ rotation curves are presented in Fig. 11.

### 3.2. Optical imaging

Optical imaging observations were performed at the Swope 40 inch telescope at Las Campanas Observatory (LCO) and the 40 inch telescope at SSO. The observations at the Swope telescope were taken with the SITe 3 chip with a pixel scale of 0.43 arcsec pixel$^{-1}$ and a field of view of 12 arcmin. The observations at SSO were taken with the direct imager and the SITe 2K detector with a pixel scale of 0.6 arcsec pixel$^{-1}$. $B$, $V$ and $I$ filters were used, three images were taken per object per band, except that two frames were obtained in some cases. Total exposure times are shown in Table 5. Three standard fields (Landolt 1992) were acquired per night. A series of twilight sky flats were obtained each night. The average seeing was ≈1.2 arcsec for the LCO runs and ≈1.3 arcsec for the SSO runs. The optical data was reduced using standard IRAF routines. Bias and overscan subtraction was performed. The bias frames at the Swope telescope showed in the last columns in the CCD reading direction an increase of 1% in the number of counts, so the bias frame was subtracted from the observations. The bias levels were very stable during the runs. Calibration for the photometric nights[1] was done using APPHOT on the Landolt fields and then using PHOTCAL to fit the zero-point magnitudes, the colour, and the extinction coefficients of the form:

$b = B + b_1 + b_2(B - V) + b_3 X$

$v = V + v_1 + v_2(B - V) + v_3 X$

$i = I + i_1 + i_2(B - V) + i_3 X$

where $B$, $V$, $I$ are the standard magnitudes, $b$, $v$, $i$ the instrumental magnitudes per second, $X$ the airmass and $b/v/i_j$ the unknown transformation coefficients. In the cases where the Thuan-Gunn $i$ filter were used, a transformation was applied from Thuan-Gunn to Cousins system using the transformation equation by Schombert et al. (1990). The resulting parameters are presented in Table 1.

### 3.3. Near-infrared imaging

NIR imaging data was obtained at Las Campanas Observatory, using the DuPont 100 inch telescope and the Swope 40 inch telescope; and also, the 2.3 m telescope at Siding Spring Observatory. The NIR observations at the DuPont were obtained with the IR camera with a Rockwell NICMOS3 HgCdTe 256×256 array with a pixel scale of 0.42 arcsec pixel$^{-1}$ (1.8 arcmin × 1.8 arcmin field of view). The IR observations at the Swope telescope were taken with the IR camera with a NICMOS3 HgCdTe 256 × 256 with a pixel scale of 0.6 arcsec pixel$^{-1}$ (2.5 arcmin × 2.5 arcmin field of view). CASPIR, the imager at the 2.3 m telescope at SSO, uses a Santa Barbara Research Center 256 × 256 InSb detector array, with a focal plane scale of 0.5 arcsec pixel$^{-1}$. The filters used were $J$, $H$, $K_n$ for the imaging with CASPIR and $J_{short}$, $H$, $K_{short}$ for the LCO runs. The NICMOS3 detector becomes nonlinear when the total counts (sky + object) exceed 17 000 ADU. For this reason and the sky variability in the NIR, we exposed for 60–120 s in $J$ 30 s in $H$ and 15 s in $K$. Each object was observed at several positions on the array. The average seeing was ≈1 arcsec for the LCO runs and ≈1.2 arcsec for the SSO runs.

---

[1] The non photometric observations at LCO were repeated afterward during photometric conditions at SSO.

Total exposure times and the observing log are presented in Table 6. For the NIR imaging obtained at LCO only the smallest angular size galaxies of the sample were observed. Therefore no offset to different sky positions was made and the sky was simply calculated by averaging a number of the stacked images for a given object with a $\sigma$ clipping rejection algorithm.

The CASPIR images were recorded at several positions on the array with a dither pattern and larger offsets were applied to obtain sky frames with the same exposure time as the object frames. The dithering between frames was typically 20 arcsec and the offset for the sky frames was typically of 2.5 arcmin. Sky frames were obtained after each second object exposure, beginning and ending each exposure sequence with a sky exposure. Each single exposure time was the same as for the Las Campanas runs.

The NIR data was reduced using standard IRAF routines and *eclipse* (IR data reduction package developed at ESO). All the frames had to be corrected for instrumental effects such as nonlinearity, dark current and pixel-to-pixel variation. The observations at LCO were carefully made in order to avoid the nonlinear regime of the array, so no correction for nonlinear deviations was needed. The response of the CASPIR detector has a quadratic nonlinearity which must be corrected before dark correction and flatfielding is done. CASPIR data is linearised by subtracting a bias frame and then applying the quadratic correction.

To better avoid the thermal emission of the telescope, series of on-lamp and off-lamp dome flats were used to flat-field the images in the case of the CASPIR data and good flat-fielding (around 1%) was achieved at LCO using twilight flats.

The NIR background has contributions from OH airglow in the $J$, $H$ and $K$ bands, moonlight, especially in the $J$ and $H$-band, and from thermal emission from the telescope and sky in the $K$ band. To construct a sky image for the LCO data, we selected a number of neighbouring images taken within ±5 min, to ensure a sky variation not larger than 1% to 2%, from the given image; then, the images are combined with a $\sigma$ clipping rejection algorithm, and finally, subtracted from the given image. For the CASPIR data we created a sky by subdividing each object observation sequence in 10 min observing blocks, creating the sky frame from the off-source images and then subtracting this sky from the on-source images of the subset. When no satisfactory subtraction was achieved the nearest off-source frames were used. Only in a few cases could the sky values be obtained from the edges of the galaxy frames for the CASPIR data (eso416-G012, IC 2822 and NGC 7246).

**Table 2.** Calibration coefficients of the NIR observations.

| Band | Zero-point | Color coeff. | Airmass coeff. |
|---|---|---|---|
| | | Du-Pont LCO | |
| $J$ | 21.55 ± 0.028 | –0.069 ± 0.08 | –0.094 ± 0.023 |
| $H$ | 21.15 ± 0.020 | –0.096 ± 0.01 | –0.056 ± 0.025 |
| $K$ | 21.11 ± 0.036 | 0.094 ± 0.11 | –0.075 ± 0.030 |
| | | March 1999, Swope LCO | |
| $J$ | 19.89 ± 0.032 | –0.082 ± 0.015 | –0.054 ± 0.020 |
| $H$ | 19.45 ± 0.054 | –0.013 ± 0.061 | –0.155 ± 0.008 |
| $K$ | 19.05 ± 0.04 | 0.030 ± 0.0256 | –0.095 ± 0.040 |
| | | May 2000, SSO | |
| $J$ | 21.35 ± 0.083 | –0.034 ± 0.053 | –0.054 ± 0.056 |
| $H$ | 21.27 ± 0.104 | –0.015 ± 0.090 | –0.152 ± 0.094 |
| $K$ | 20.02 ± 0.15 | 0.092 ± 0.073 | –0.20 ± 0.12 |
| | | August 2000, SSO | |
| $J$ | 21.42 ± 0.098 | –0.027 ± 0.062 | –0.031 ± 0.042 |
| $H$ | 21.15 ± 0.068 | –0.013 ± 0.020 | –0.08 ± 0.03 |
| $K$ | 20.41 ± 0.069 | 0.075 ± 0.064 | –0.27 ± 0.243 |

For flux calibration a number of standard stars from the HST standards list Persson et al. (1998) and the IRIS standard star list Carter (1995) were used. These lists provide stars fainter than 8th magnitude to avoid saturation of the detector during the acquisition process. 6 to 7 standard star observations were done per night, whenever possible repeating each standard up to 3 times. When comparing different photometric systems, transformation equations can be derived comparing common standard stars. To transform from the MSSSO system to the CTIO system the equations derived by McGregor (1994) were used.

The difference between $K_n$ and $K_{short}$ is minimal, giving a very similar band pass. The transformation between $K$ and $K_n$ was determined from model stellar atmosphere energy distributions and measured filter transmission curves (McGregor 1995, CASPIR manual):

$$K_n = K - 0.022(J - K).$$

The NIR calibration coefficients for the different runs are presented in Table 2.

## 4. Preparing the images for the modelling

### 4.1. Star removal techniques

In order to model the gas flows using SPH, a smooth mass distribution map has to be input in the code. From the observed light distribution one can derive a mass model. However, the foreground stars in the field of the galaxies have to be removed so that the light from these sources does not affect the mass distribution and the SPH modelling. Furthermore, the ellipse fitting applied to the galaxies would also be affected by the light of these foreground stars. For the non-saturated stars, a point spread function (PSF) was calculated for each frame. The stars in each frame were found by using a detection algorithm which searches for the local density maxima with a given FWHM and a peak intensity greater than certain value. Then, photometry is done on the found stars; and finally, the scaled PSF centered on each object is subtracted. In some cases, the residuals had to be cleaned using a linear interpolation algorithm. In the case of saturated stars, a model for each of the stars was created using a King profile for the halo and the star itself for the center. In most of the cases, the diffraction spikes were not cleanly removed and linear interpolation had to be done. For the finding stars algorithm, photometry and PSF matching, routines within DAOPHOT in IRAF were used. In the case where small background galaxies appeared, a box enclosing the galaxy was replaced by the pixel values of a box of the same size located at the same radial position at 180°. Codes in IDL were written for the linear interpolation and the saturated star modelling. An example of the cleaning procedure is shown in Fig. 3.

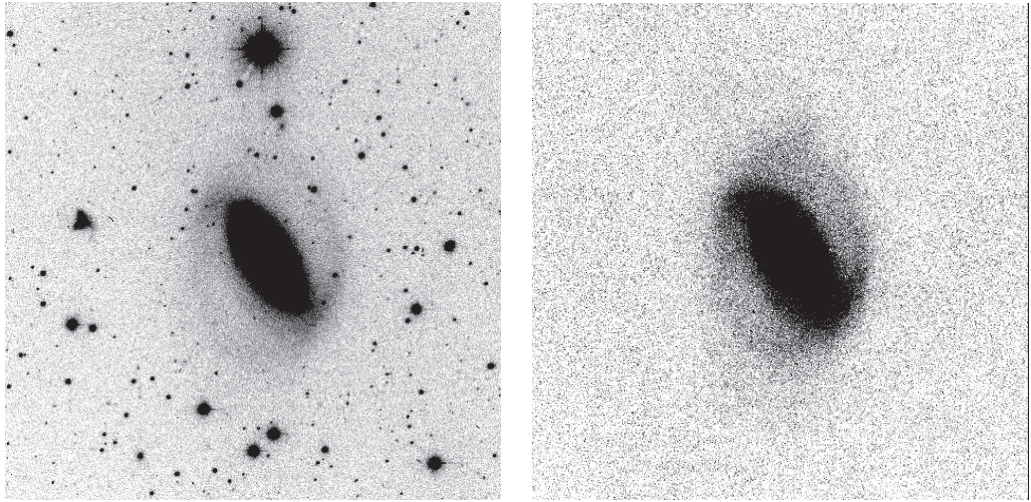

**Fig. 3.** NGC 5728 *I*-band before and after star removal.

## 4.2. Combining H-band and I-band images

To model the stellar mass distribution, we would ideally prefer to use a photometric band that is not affected by dust. The bands that offer a good compromise between representing the old stellar population and being less affected by dust, the NIR bands, are nevertheless very time consuming to map due to the small size of the NIR arrays, to the fact that they are not as sensitive as the optical CCDs to obtain deep large images, and to the sky brightness limitation for NIR imaging.

But, are galaxy disks optically thick all throughout their disks or are they optically thin in the outer regions and opaque or moderately optically thick in their inner parts?. Recent studies by Xilouris et al. (1999) present evidence showing that disks are optically thick in their inner parts and transparent in the outer parts in the *I*-band, for the inclinations encountered in this work (between 30° and 70°). Therefore for our sample it is a good approach to combine NIR data for the inner parts of galaxies with *I*-band data for the outer parts.

In order to obtain the mass distribution from the light down to the fainter isophotes, the *I*-band has been used for the outer optically thin regions while the *H*-band has been used for the inner regions of the disk. To combine both images first we have to create images of the same dimensions. Then, both frames are registered using common stars and making use of the GEOMAP task within IRAF to create the transformation function. GEOMAP determines higher order coordinate transformations (coordinates, rotation, pixel scale). GREGISTER transforms images from one coordinate system to another, applying the transformation map created by GEOMAP; then resampling was done using linear interpolation. Once the images have been centered and have the same pixel scale, the PSF of the two images are matched by convolving with a kernel computed directly from objects in the image with worst seeing. If $f$ is the reference image, $h$ is the best seeing image and $k$ is the PSF matching function then by the convolution theorem;

$$f = k * h,$$

$$\tilde{k} = \frac{\tilde{f}}{\tilde{h}}$$

where $\sim$ denotes the Fourier transform. Once the images have been matched, a mask is created to isolate in both images the regions that will be combined. These regions are chosen using the extinction values obtained from the spectroscopy and choosing the *H*-band region to extend into the optically thin regions ($\tau \approx 1$) where the observed colours do not change significantly. The noise in the *H*-band images is higher than in the *I*-band images. The SPH code deals with the different noise levels in the different parts of the images.

Before the images were finally combined, they were also deprojected. This deprojection is done assuming a flat disk and then simple geometrical stretching is done. An example is shown in Figs. 4 and 5, where the masked images of NGC 7267 and the respective central line profiles are presented. One can note the fact that the *H*-band seems to be more symmetric and smoother; this is more evident if one looks at the combined image and the *I*-band image, due to the higher dust sensitivity in the *I*-band compared to the *H*-band.

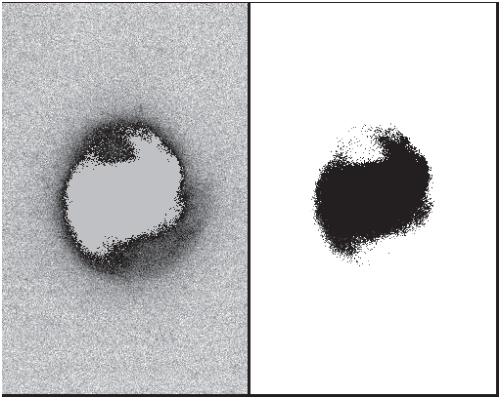

**Fig. 4.** On the left the *I*-band image of NGC 7267 masked in the inner parts. On the right, the *H*-band image of the same galaxy masked in the outer parts. Both masks are used on their respective images and then combine into a single light distribution.

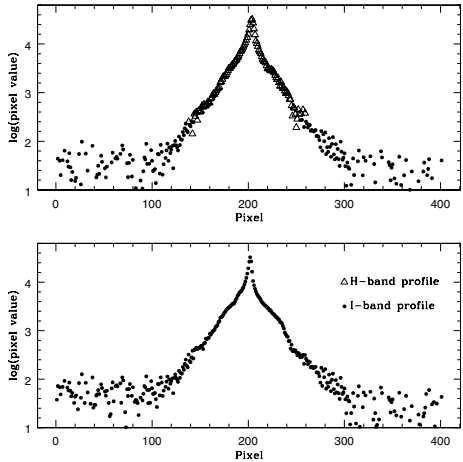

**Fig. 5.** Profiles of a cut along the major axis of NGC 7267 for the combined profile (*top panel*) and the *I*-band profile (*lower panel*). Notice the smooth transition in the combined image from the *H*-band region to the *I*-band region.

## 5. General galaxy parameters

To quantitatively describe the sample, some global parameters were extracted from the data. The total magnitude in the different bands were derived by integrating the flux within a certain radius corresponding to the point where the isophotal magnitude is 25 mag arcsec$^{-2}$ in the *B*-band (see Table 7). For the galaxies for which the isophotes do not reach 25 mag arcsec$^{-2}$ in the *B*-band, the surface brightness was extrapolated to the required magnitude using the exponential disk parameters obtained from the surface brightness profiles (Sect. 5.1). The total magnitudes have been checked both internally (measurements of the same galaxy at different runs and telescopes) and externally (measurements of the same galaxy by different authors). For the internal comparison the difference in *H* magnitude obtained at the 40 inch telescope at LCO and the magnitude obtained at the 40 inch telescope at SSO is well within the errors (11.25 ± 0.10 and 11.07 ± 0.02 respectively for UGC 10384 and 10.42 ± 0.06 and 10.32 ± 0.01 for NGC 7483). Two of the sample galaxies (NGC 6012 and NGC 5728) have integrated *K* and *H* magnitudes from previous works (Márquez & Moles 1999; de Jong 1996a; Glass & Moorwood 1985). These results are also in good agreement with the results obtained in this study.

*I*-band contours, surface brightness profiles in all available bands, rotation curves and a plot of the ellipticity and the position angle are presented in Fig. 11. At the end of the section the methodology and results for the extraction of the parameters of the non-axisymmetric structures will be presented. In the inner parts the ellipticity and the position angle (PA) obtained from the ellipse fitting technique will be the indicators of the existence of non-axisymmetric structures such as bars and the position and size of these structures (Márquez et al. 1999, 2000).

The velocity used in Sect. 2 is the asymptotic velocity obtained from fitting a parametric rotation curve to the observed one. The formula representing a galaxy by a sequence of flattened spheroids by Brandt (1960) and by Bettoni & Galetta (1997) was adopted, namely;

$$v(r, \phi) = 3V_{\max} \frac{r/r_{\max}}{1 + 2(r/r_{\max})^{3/2}} P(i, \text{PA} - \phi)$$

where, $V_{\max}$ is the maximum velocity and the radius $r_{\max}$ is the radius at which this velocity is reached, $i$ is the inclination, PA is the position angle of the line of nodes, $\Phi$ is the observed position angle and $P(i, \text{PA} - \phi)$ is the projection factor. Kannappan et al. (2002) derived a velocity value by fitting the empirical function:

$$v = v_0 + v_c \frac{(1 + x)^\beta}{(1 + x^\gamma)^{1/\gamma}}$$

where $x = r_t/(r - r_0)$, $(r_0, v_0)$ define the origin, $v_c$ gives a velocity scale and $r_t$ is related to the turnover radius. $\beta$ and $\gamma$ are free parameters in the fit.

To check the difference between the asymptotic velocities derived by the two parametric functions we did the following test: We used the on-line data catalog by Courteau (1997), which uses the same analytical function as Kannappan to derived the velocities, to construct artificial rotation curves. These rotation curves were then used as an input to the fit of the flattened spheroids. Then, the derived parameters were compared, finding a difference with the fitted values in the asymptotic velocity of $\approx \pm 8$ km s$^{-1}$. The $V_{\max}$ and $r_{\max}$ are presented in Table 8. The final error in the $V_{\max}$ is of the order of 15 km s$^{-1}$ and the error in $r_{\max}$ is of the order of 10 arcsec.

### 5.1. Surface photometry

To obtain the surface density profiles, first the contribution from the background has to be carefully removed. This was done by taking the mean value of boxes placed around the galaxies in regions far from the target galaxies. The standard deviation of the median of the distribution in the different boxes was adopted as the error in the background determination. The average azimuthal profiles were obtained by fitting elliptical isophotes to the galaxy frames using the task ELLIPSE in IRAF. The distances used to evaluate the luminosities were derived from the measured redshift corrected for galactocentric motion.

The photometric decomposition of the light distribution was carried out by 1-D fitting of the surface brightness profiles in all the bands with an exponential law for the disk (Freeman 1970) and the $r^{1/4}$ law for the bulge (de Vaucouleurs et al. 1991). The bulge effective radius and the disk scale-length were obtained using the "marking the disk" method, which consists of fitting the linear part of the luminosity profile, when plotted on a magnitude scale, and then fitting the power law. In this case the region containing the bar was not used to fit the linear part. The results of the fitting parameters are presented in Table 8. The effect of using the "marking the disk" method in the disk parameters has been widely studied (Knapen & van der Kruit 1991), they showed that the uncertainties using this method are of the order of 15%. The situation is different for the bulge parameters, a $r^{1/4}$ law is not necessarily the best fit to the bulge light distribution (Andredakis et al. 1995; de Jong 1996b); the difference in the parameters from fitting a $r^{1/4}$ to fitting an exponential bulge is about 50% for the later types and around 25% for the earlier types in the sample galaxies.

### 5.1.1. Comparison of the results with other works

It is difficult to compare 1-D to 2-D decomposition techniques from different authors, since there are inconsistencies in how the ellipses are fit for 1-D and 2-D techniques (Moriondo 1997); also, in some cases bars are not taken into account for the 1-D fit while they are introduced in the 2-D fit (de Jong 1996b).

The relation between the parameters of the disk and the bulge are given in Figs. 6 and 7. The resulting parameters are compared to the data presented by de Jong (1994) of a sample of non perturbed and non-peculiar galaxies, to the data by Márquez & Moles (1999) of a sample of isolated galaxies, and to the data by Baggett et al. (1998) of a large sample of spiral galaxies. The parameter extraction in the works by Márquez & Moles (1999) and by Baggett et al. (1998) are very similar to the methodology presented here; therefore, the comparison is more meaningful than comparing the results to the data by de Jong (1994), where it is used a 2-D decomposition technique. However, it is still interesting to plot the range covered by de Jong (1994) together with the results from this and previous works. Surface brightness parameters and scale parameters seem to be correlated, in agreement with previous works. Regarding the Kormendy relation (Kormendy 1977), we find that contrary to (Márquez & Moles 1999) and (Baggett et al. 1998), and in agreement with (de Jong 1996a), no correlation between $\mu_e$ and $r_e$ is found in this data, possibly due to the small range in bulge magnitudes in our sample and de Jong's sample. No significant differences are found among the properties of field and isolated galaxies although they display tighter parameter relations compared to interacting galaxies, in agreement with Márquez (see right panels of Figs. 6 and 7).

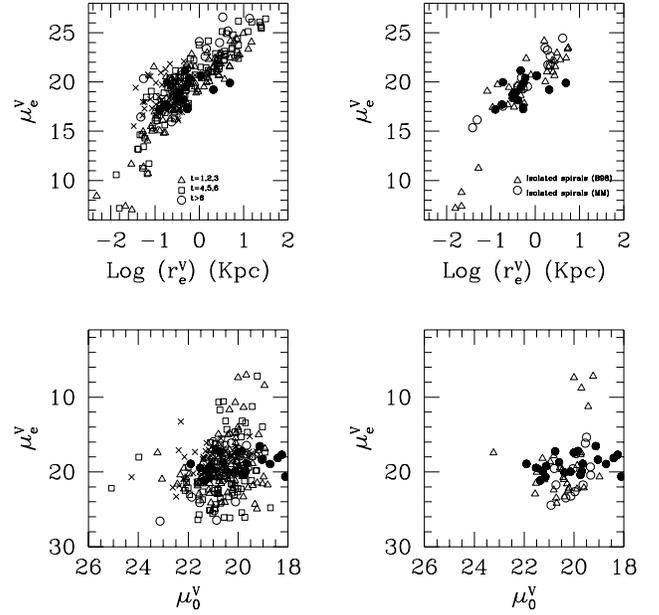

**Fig. 6.** Relations for the bulge and disk parameters. $\mu$ is the surface brightness, $h$ the disk scalelength and $r_e$ the effective radius. The left panels shows the data from all the spirals from Baggett et al. (1998) (Sa to Sb are plotted as open triangles, Sbc to Scd as open squares and later types than Scd are plotted as open circles). The crosses are the galaxies from de Jong (1994) and the full black dots are the galaxies from this work. In the right panels only the isolated galaxies are plotted. The open circles is the data from Márquez & Moles (1999), the open triangles are the galaxies from Baggett et al. selected as isolated and the full black dots are the galaxies from this work.

### 5.2. Tully-Fisher relation for the sample galaxies

Figure 8 shows the Tully-Fisher (T-F) relation for all the galaxies of the sample. The distance was calculated from the redshift assuming a $H_0$ of 75 km s$^{-1}$ Mpc$^{-1}$. The $B$ and $H$-band correspond to the observed corrected integrated magnitude and the maximum rotational velocity is obtained from the observed optical rotation curves as indicated in Sect. 5. In order to compare the data from our sample with T-F from data in literature an equation had to be chosen for consistency. The relation fitted to the data was of the form:

Magnitude$_{band}$ = Zeropoint + slope(log $W_{20}$ − 2.5).

A correction was applied to the maximum optical rotational velocity (observed $V_{max}$) in order to convert it to the 20% line width ($W_{20}$) scale used in Sakai et al. (2000). First, the conversion from optical to the 50% line width ($W_{50}$) used by Kannappan et al. (2002) was done. This conversion was calculated empirically from iterative least-squares fit to a sample data.

$W_{50} = 19(\pm 6) + 0.90(\pm 0.03)(2V_{max})$.

To adjust for the difference between $W_{50}$ and $W_{20}$ line-widths, 20 km s$^{-1}$ was added to $W_{50}$. This number is derived from a comparison between both measures done by Kannappan;

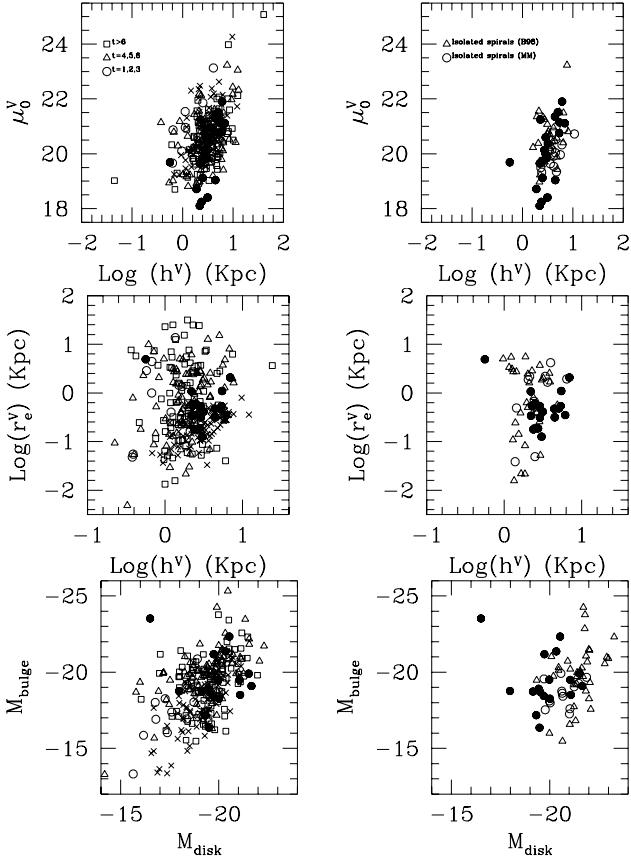

**Fig. 7.** Relations for the bulge and disk parameters. Coding is the same as in Fig. 6. Notice the general agreement between the different samples. The figure shows a clear correlation between the surface brightness parameters and the scale parameters. No correlation is found for the effective radius of the bulge against the scale-length of the disk, although there seems to be some correlation for the isolated sample. The lower two panels show a clear correlation between the magnitudes of the disk and the bulge.

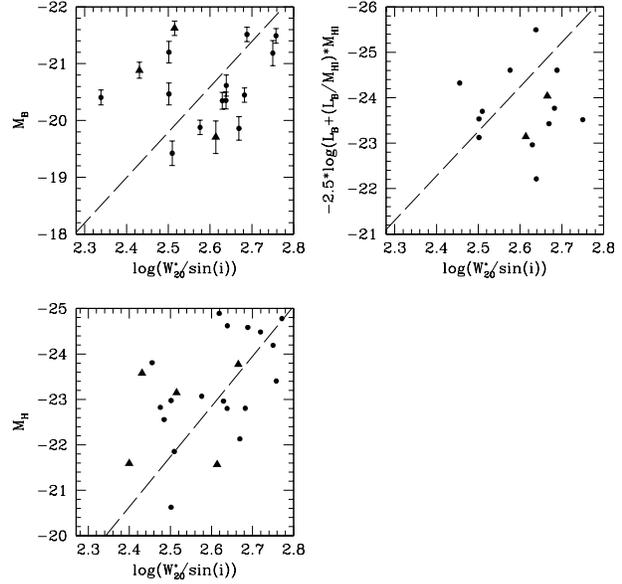

**Fig. 8.** Tully Fisher relation for the sample galaxies. The top left panel shows the Tully-Fisher relation for the *B*-band luminosity and the log $W_{20}$ velocity width. The errors in the y axis represent the dispersion due to an average peculiar velocity if 300 km s$^{-1}$ added quadratically to the T-F dispersion presented in Sakai et al. (2000). The right panel shows the baryonic T-F relation for the sample galaxies which had HI measurements in literature. The lower panel shows the T-F for the sample galaxies in the *H*-band. The dash line corresponds to the Tully-Fisher relation determined by Sakai et al. (2000). The full dots represent the galaxies with inclinations lower than 60° and the triangles galaxies with inclination greater or equal to 60°. Notice that the dispersion in the relation decreases for the *H*-band and the baryonic T-F cases.

Haynes et al. (1999) derived a similar number from a different sample. The integrated magnitudes were corrected for inclination and galactic extinction. The dashed line (Fig. 8) corresponds to the Tully-Fisher relation determined by Sakai et al. The slope of the *B*-band is $-7.97 \pm 0.72$ and the zero point $-19.80 \pm 0.11$. The same slope was found for the *baryonic* relationship and a slope of $-11.03 \pm 0.87$. For the *H*-band data a zero point of $-21.74 \pm 0.86$ was found.

The S0 galaxy NGC 6014 was not included in the T-F figures due to the fact that it lies well outside the diagram with log $W_{20} = 2.867$ and $M_B = -19.38$, obviously the T-F relation cannot be applied in this case of a lenticular galaxy.

There is no significant difference in dispersion between the different bands although the NIR and the *baryonic* relationship seem to be tighter.

### 5.3. Bar parameters determination

The rotation curve of an axisymmetric galaxy does not provide enough information to distinguish between the contribution to the mass distribution from the halo and the disk. The non circular flows in the bar potential can help to disentangle this degeneracy. Barred galaxy models can be characterised with four parameters: the central concentration, the axial ratio, pattern speed and the relative mass of the bar. These parameters will affect the bar potential and therefore the streaming motions and the gas shocks in the bar region. Both pattern speed and central concentration affect the rotation curve and the position of resonances.

We want to describe the bars of the galaxies to be modelled in order to use these parameters for the simulations and to characterise the modelled galaxies. We can do a preliminary characterisation of the barred galaxies of the sample by extracting one of the four main parameters which is accessible directly from the images; the axial ratio. However, defining the ends of the bars is not an easy task. The bar morphology depends on the wavelength chosen and deprojection uncertainties alter the observed axis ratio. As shown by numerical simulations (Kalnajs, private communication), for a self gravitating disk with no dark halo there would be a minimum axis ratio (determined by stability criteria) of 0.4. Athanassoula (2002) shows that bars grown in a more halo-dominated environment are stronger than bars grown in disk-dominated environments due to the transfer of angular momentum between the two components. Therefore, the bar axis ratio could test in a morphological way the need for a massive dark halo in the inner parts and check the maximum disk hypothesis.

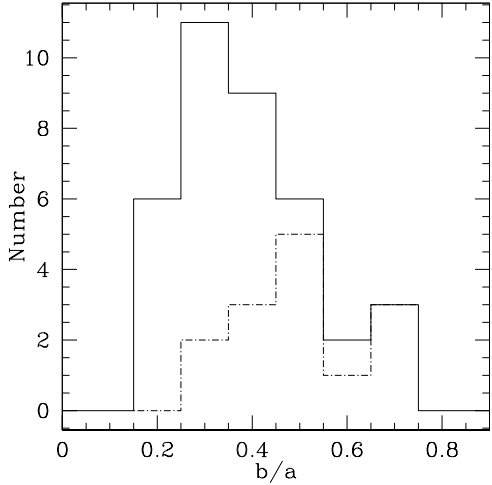

**Fig. 9.** Distribution of bar axis ratios for the barred sample galaxies and de Jong's data. The full line corresponds to all the data together and the dashed line to the sample presented here. de Jong axis ratios are systematically lower than the ones found for us. Both samples use the NIR bands to calculate the bar axis ratios.

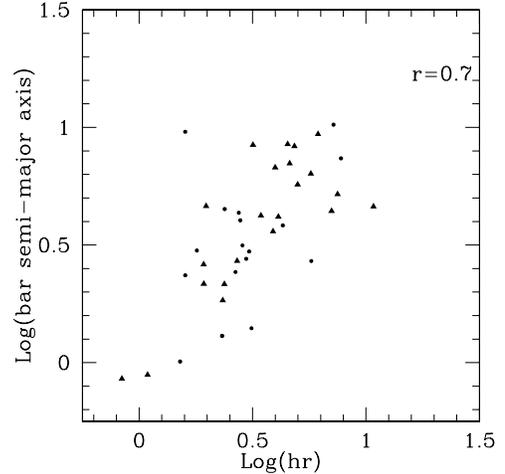

**Fig. 10.** Bar size vs disk scale-length for the barred sample galaxies plus de Jong's data. The triangles represent the barred sample galaxies and the full dots de Jong's data. Notice the correlation coefficient between the scale-length and the bar's semi-major axis. Possibly due to the rapid growth of the bar.

There are different techniques for extracting the bar size: (i) visually; (ii) computing Fourier amplitudes from the ellipse fitting; (iii) examining the behaviour of the ellipticity and position angle as a function of radius for barred isophotes defining the end of the bar as the radius where the PA sharply changes at the same radius as the ellipticity decreases to reach a local minimum, Wozniak et al. (1995); (iv) fitting an analytical function. The Fourier amplitudes method suffers from the errors of the ellipse fitting and the fact that one is dealing with second order effects obtained from noisy data. The behaviour of the ellipticity and position angle also suffers from the ellipse fitting errors, giving probably a slightly underestimated size of the bar. In this work, the isophotal analysis technique was used to extract the bar parameters. The axis ratio extraction was done from the $H$-band images which should be less affected by dust and a better indicator of the stellar mass distribution. The distribution of the bar axis ratios for the galaxies of this sample and de Jong's sample is presented in Fig. 9. The difference in the methodology used to extract the major and minor axis of the bar; and most of all, the low number statistics makes very difficult to draw any conclusions about the bar axis ratios and maximum disk. It would be very interesting to extract the bar axis ratio for a much larger sample of NIR images in a consistent way to explore the statistical range spanned by the data and compare it to the models.

No correlation with morphological type has been found in the axis ratios. However, Fig. 10 shows a clear correlation between the bar semi-major axis and the disk scale-length of the disk. The absence of small bars in large disks could be related to the fact that bars form quickly and statistically what we see are only fully formed bars. Recent simulations of bar formation in galaxies with dark matter halos (Valenzuela & Klypin 2003) find that the bar semi-major axis is close to the exponential length of the disk, in agreement with our observations.

## 6. Conclusions

We have presented the surface photometry and ionised gas rotation curves for a sample of 27 isolated spiral galaxies. Although the number of active nuclei in the sample is small (4 out of the 27 galaxies), the active galaxies show similar properties to the non-active galaxies in agreement with Márquez et al. (2004). We have re-classified as a Sy2 active nuclei a galaxy previously lacking nuclear classification (NGC 6014). The ellipse fitting analysis and visual check showed that, from 8 galaxies with no bar information, 3 presented clearly in the NIR or the $I$-band a bar morphology (UGC 10130, UGC 4861 and UGC 1553). From the 3 galaxies previously classified as non-barred, from optical images, two showed bar morphologies in the NIR (NGC 6014 and ESO 417-G006).

We have presented the methodology used to clean and create the composite images that served to model the stellar mass distribution of the galaxies used to model the velocity fields (Paper I).

The relations derived for the disk and bulge parameters show that the surface brightness and scale parameters are correlated, in agreement with previous works (de Jong 1996a; Márquez & Moles 1999; Baggett et al. 1998). Contrary to Márquez & Moles and Baggett et al., and in agreement with de Jong, no correlation between $\mu_e$ and $r_e$ is found in this data, possibly due to the small range in bulge magnitudes in our and de Jong's sample.

With respect to the influence of the environment on the derived parameters; when comparing the derived bulge and disk parameters of our sample of isolated galaxies with other studies of isolated and field galaxies, no significant differences are found among the properties of field and isolated galaxies. Although, isolated galaxies display tighter parameter relations. This result is in agreement with the conclusions by Márquez et al. (1999).

The T-F relation derived for these galaxies is similar to that of galaxies in denser environments. There is no significant difference in dispersion between the relation derived in different bands; although, the baryonic relationship seems to be slightly tighter than the T-F derived for the photometric bands alone.

We find a correlation between the bar size and the disk scale-length for the barred galaxies of the sample. This might be reflecting the expected rapid growth of the bar. However, the sample is biased toward early-type bars and a bigger sample would be needed to conclude whether this correlation reflects the bar evolution.

*Acknowledgements.* This work was supported by the Australian National University through an ANU PhD. grant. We specially acknowledge and thank J. Maza for his help and useful suggestions. We want to thank the anonymous referee for the useful comments; which helped to improve this manuscript, and to R. Peletier for the useful discussions. I.M. acknowledges financial support from the Junta de Andalucía and DGIyT grants AYA2001-2089 and AYA2003-00128.